\documentclass[
    ,final            
  ]
  {aipproc}

\layoutstyle{6x9}

\usepackage{graphicx}
\usepackage{amsmath}
\usepackage{color}

\definecolor{darkgreen}{rgb}{0,0.5,0}
\definecolor{purple}{rgb}{0.5,0,0.5}
\definecolor{nblue}{rgb}{0.0,0.0,0.50}
\definecolor{scarlet}{rgb}{1.0,0.2,0}

\newcommand{\lsim}{\mathrel{\rlap{\lower4pt\hbox{\hskip0pt$\sim$}}
\raise1pt\hbox{$<$}}}           
\newcommand{\gsim}{\mathrel{\rlap{\lower4pt\hbox{\hskip0pt$\sim$}}
\raise1pt\hbox{$>$}}}           

\begin{document}

\title{Dressed-quarks and the Roper resonance}


\author{C.\,D.~Roberts,\!$^{1,2,3,4}$ I.\,C.~Clo\"et,\!$^5$ L.~Chang$^1$ and H.\,L.\,L.~Roberts$^6$}
{address={
  $^1$Physics Division, Argonne National Laboratory, Argonne, Illinois 60439, USA\\
  $^2$Institut f\"ur Kernphysik, Forschungszentrum J\"ulich, D-52425 J\"ulich, Germany\\
  $^3$Department of Physics, Illinois Institute of Technology, Chicago, Illinois 60616-3793, USA\\
  $^4$Department of Physics, Center for High Energy Physics and State Key Laboratory of Nuclear Physics and Technology, Peking University, Beijing 100871, China\\
  $^5$Department of Physics, University of Washington, Seattle WA 98195, USA\\
  $^6$Physics Department, University of California, Berkeley, California 94720, USA
  }}

\begin{abstract}
A Dyson-Schwinger equation calculation of the light hadron spectrum, which correlates the masses of meson and baryon ground- and excited-states within a single framework, produces a description of the Roper resonance that corresponds closely with conclusions drawn recently by EBAC.  Namely, the Roper is a particular type of radial excitation of the nucleon's dressed-quark core augmented by a material meson cloud component.  There are, in addition, some surprises.
\end{abstract}

\maketitle

\begin{figure}[t]
\includegraphics[clip,width=0.66\textwidth]{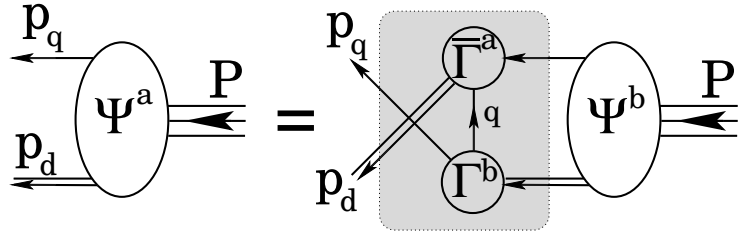}
\caption{\label{faddeevfigure} Poincar\'e covariant Faddeev equation, employed in Ref.\,\protect\cite{Roberts:2011cf} to calculate the properties of nucleon and $\Delta$ ground- and excited-states.  $\Psi$ is the Faddeev amplitude for a baryon of total momentum $P= p_q + p_d$.  It expresses the: relative-momentum correlation between the dressed-quark and -diquarks within the bound state; and contributing strength of admissible diquark correlations.  The shaded region demarcates the kernel of the Faddeev equation, in which: the \emph{single line} denotes a dressed-quark propagator; $\Gamma$ is a diquark Bethe-Salpeter-like amplitude; and the \emph{double line} is a diquark propagator.}
\end{figure}

\hspace*{-\parindent}\mbox{\textbf{1.~Introduction}}~~No approach to QCD is comprehensive if it cannot provide a unified explanation of both mesons and baryons.  Dynamical chiral symmetry breaking (DCSB) is a fact in QCD \cite{CDRobertsI}.  It has an enormous impact on meson properties \cite{Chang:2011vu} and must also be included in the description and prediction of baryon properties.  The Dyson-Schwinger equations (DSEs) furnish the only extant framework that can simultaneously connect both meson and baryon observables with this basic feature of QCD \cite{Eichmann:2008ef}.  DCSB is an essentially quantum field theoretical effect.  In quantum field theory a meson appears as a pole in the four-point quark-antiquark Green function.  From this observation one can derive the Bethe-Salpeter equation.  It is therefore unsurprising that a baryon appears as a pole in a six-point quark Green function; and one can derive a Poincar\'e covariant Faddeev equation to describe the associated bound state \cite{Cahill:1988dx}.  The Faddeev equation sums all possible exchanges and interactions that can take place between three dressed-quarks.  These dressed-quarks are one of the basic consequences of DCSB in QCD.  A tractable Faddeev equation, Fig.\,\ref{faddeevfigure}, follows from the observation that an interaction which describes color-singlet mesons also generates quark-quark (diquark) correlations in the color-antitriplet channel \cite{Cahill:1987qr}.

It should be emphasized that the diquark correlations within baryons are nonpointlike.  They have a nonzero extent, which can be characterized by a charge radius. One finds, e.g., that isoscalar-scalar diquark correlations have a charge radius commensurate with that of the pion and the analogous axial-vector diquarks have radii similar to those of the $\rho$-meson \cite{Roberts:2011wy}.  Quantum mechanical models that employ pointlike diquark degrees of freedom have no relation to the Faddeev equation description of baryons in quantum field theory.

\begin{figure}[t]
\hspace*{2em}\begin{minipage}[t]{1.05\textwidth}
\begin{minipage}[t]{\textwidth}
\includegraphics[clip,width=0.95\textwidth]{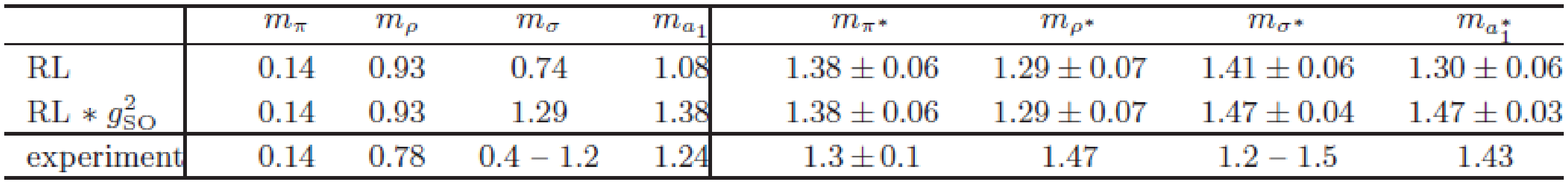}\\
\end{minipage}
\begin{minipage}[t]{\textwidth}
\includegraphics[clip,width=0.95\textwidth]{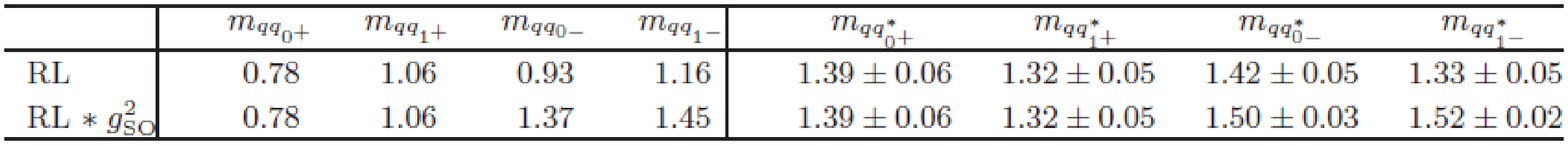}
\end{minipage}
\end{minipage}
\caption{\label{masses2}
\emph{Upper panel} -- Meson masses (GeV) computed using a contact-interaction DSE kernel, which produces a momentum-independent dressed-quark mass $M=0.37\,$GeV from a current-quark mass of $m=7\,$MeV.  ``RL'' denotes rainbow-ladder truncation.  Row-2 is obtained by augmenting the RL kernel with repulsion generated by vertex dressing \protect\cite{Chang:2011ei}.  Row-3 lists experimental masses \protect\cite{Nakamura:2010zzi} for comparison.  NB.\ Isospin symmetry is implemented so, e.g., $m_\omega=m_\rho$, $m_{f_1}=m_{a_1}$, etc.
\emph{Lower panel} -- Diquark masses computed using the same setup.  The states are labeled by their $J^P$.
In both panels, superscript-$\ast$ denotes a radial excitation.
}
\end{figure}

\hspace*{-\parindent}\mbox{\textbf{2.~Faddeev Equation}}~~The kernel of the Faddeev equation is known once one has calculated the dressed-quark propagator, diquark Bethe-Salpeter amplitudes and diquark propagators.  These elements were computed in Ref.\,\cite{Roberts:2011cf} using a confining and symmetry-preserving regularization of a vector$\,\times\,$vector current-current interaction and working primarily within the rainbow-ladder truncation of the associated Dyson-Schwinger equations, which is the leading-order in a systematic, symmetry-preserving scheme \cite{Bender:1996bb}.  However, lessons learnt from more sophisticated studies of mesons \cite{Chang:2011ei} were used to improve the description of the parity-partners of the $\pi$- and $\rho$-mesons, and hence the partners of the scalar and axial-vector diquarks.  The spectra thus obtained are presented in Fig.\,\ref{masses2}.
Particular highlights are: the result for the scalar meson ground state, which matches an estimate for the $\bar q q$-component of the $\sigma$-meson obtained using unitarized chiral perturbation theory \cite{Pelaez:2006nj}; and first predictions for the masses of diquark correlations relevant to other than nucleon and $\Delta$ ground-states.

A particularly simple kernel for the Faddeev equation is obtained by employing a variant of the so-called ``static approximation,'' which itself was introduced in Ref.\,\cite{Buck:1992wz} and has subsequently been used in studies of a range of nucleon properties \cite{Bentz:2007zs}.  This is an internally consistent treatment of the contact interaction.  It leads to baryon Faddeev amplitudes that are independent of relative momentum and should produce reliable masses for those states whose rest-frame amplitudes are not dominated by components with large quark orbital angular momentum.

\bigskip

\begin{figure}[t]
\begin{minipage}[t]{1.0\textwidth}
\begin{minipage}[t]{\textwidth}
\includegraphics[clip,width=1.0\textwidth]{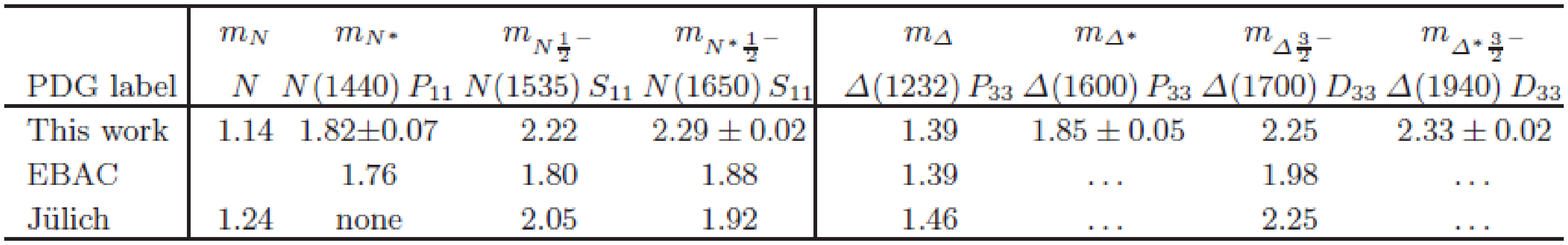}
\end{minipage}
\begin{minipage}[t]{\textwidth}
\includegraphics[clip,width=1.0\textwidth]{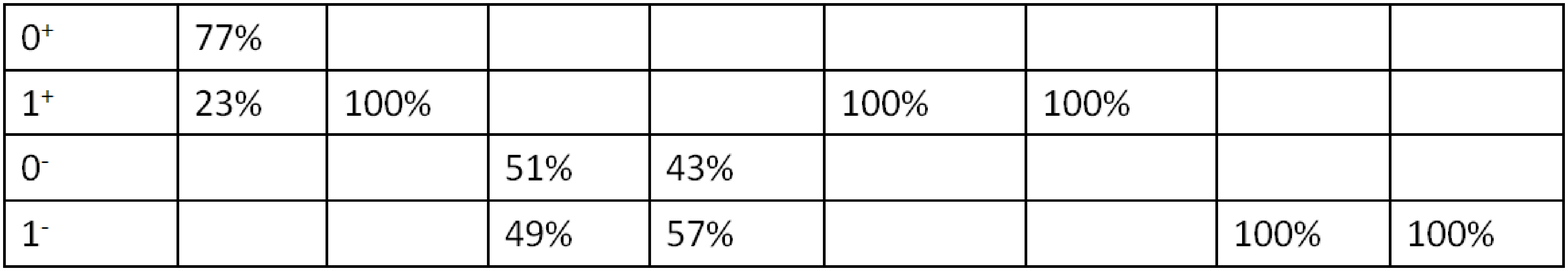}
\end{minipage}
\end{minipage}
\caption{\label{massesN}
\emph{Upper panel} --
Dressed-quark-core masses (GeV) for nucleon and $\Delta$, their first radial excitations (denoted by ``$\ast$''), and the parity-partners of these states. 
\emph{Row-2}: Bare-masses inferred from a coupled-channels analysis at the Excited Baryon Analysis Center (EBAC) \protect\cite{Suzuki:2009nj}.  EBAC's method does not produce a bare nucleon mass.
\emph{Row-3}: Bare masses inferred from the coupled-channels analysis described in Ref.\,\protect\cite{Gasparyan:2003fp}, which describes the Roper resonance as dynamically-generated.
In both these rows, ``{\ldots}'' indicates a state not found in the associated analysis.
A visual comparison of these results is presented in Ref.\,\protect\cite{CDRobertsI}.
\emph{Lower panel} -- Diquark content of the baryons' dressed-quark cores.
}
\end{figure}

\hspace*{-\parindent}\mbox{\textbf{3.~Roper Resonance}}~~Computed results \cite{Roberts:2011cf} for the dressed-quark-core masses of the nucleon and $\Delta$, their first radial excitations (denoted by ``$\ast$''), and the parity-partners of these states are presented in Fig.\,\ref{massesN}.  With these results one has simultaneously correlated the masses of meson and baryon ground- and excited-states within a single framework.  The masses should not be compared directly with experiment because the kernels employed in their calculation do not incorporate the effect of meson loops.  However, a fair comparison may be made with bare-masses inferred from sophisticated coupled-channels analyses of $\pi N$ scattering data up to $W\lsim 2\,$GeV \cite{Suzuki:2009nj,Gasparyan:2003fp}.

The predictions for the baryons' dressed-quark-cores match the bare-masses determined in Ref.\,\cite{Gasparyan:2003fp} with a root-mean-square (rms) relative-error of 10\%.  Notably, however, Ref.\,\cite{Roberts:2011cf} finds a quark-core to the Roper resonance, whereas within the J\"ulich coupled-channels model this structure in the $P_{11}$ partial wave is unconnected with a bare three-quark state.  In connection with EBAC's analysis, the predictions for the bare-masses agree within a rms relative-error of 14\%.  Notably, EBAC \cite{Suzuki:2009nj} does find a dressed-quark-core for the Roper resonance, at a mass which agrees with the Faddeev equation prediction.  As described in Ref.\,\cite{CDRobertsI}, this provides finally for an understanding of the Roper resonance.

Additional analysis within the framework of Ref.\,\cite{Roberts:2011cf} suggests a fascinating new feature of the Roper, which is evident in the lower panel of Fig.\,\ref{massesN}.  The nucleon ground state is dominated by the scalar diquark, with a significantly smaller but nevertheless important axial-vector diquark component.  This feature persists in solutions obtained with more sophisticated Faddeev equation kernels (see, e.g., Table~2 in Ref.\,\cite{Cloet:2008re}).  From the perspective of the nucleon's parity partner and its radial excitation, the scalar diquark component of the ground-state nucleon actually appears to be unnaturally large.

One can nevertheless understand the structure of the nucleon.  As with so much else, the composition of the nucleon is intimately connected with DCSB.  In a two-color version of QCD, the scalar diquark is a Goldstone mode, just like the pion \cite{Roberts:1996jx}.  (This is a long-known result of Pauli-G\"ursey symmetry.)  A ``memory'' of this persists in the three-color theory and is evident in many ways.  Amongst them, through a large value of the canonically normalized Bethe-Salpeter amplitude and hence a strong quark$+$quark$-$diquark coupling within the nucleon.  (A qualitatively identical effect explains the large value of the $\pi N$ coupling constant.) There is no such enhancement mechanism associated with the axial-vector diquark.  Therefore the scalar diquark dominates the nucleon.
The effect on the Roper is striking, with orthogonality of the ground- and excited-states forcing the Roper to be constituted almost entirely from the axial-vector diquark correlation.  One may reasonably expect this to have a material impact on the momentum-dependence of the nucleon-to-Roper transition form factor \cite{ICCloetI}.



\bigskip

\hspace*{-\parindent}\mbox{\textbf{Acknowledgments}}~~We acknowledge valuable discussions with S.\,M.~Schmidt and financial support from the Workshop.
Work supported by:
Forschungszentrum J\"ulich GmbH;
and U.\,S.\ Department of Energy, Office of Nuclear Physics, contract nos.~DE-FG03-97ER4014 and DE-AC02-06CH11357.

\vspace*{-2ex}



\bibliographystyle{aipproc}   


\begin{thebibliography}{99}

\bibitem{CDRobertsI} C.\,D.~Roberts, ``Opportunities and Challenges for Theory in the N$^\ast$ program,'' these proceedings and arXiv:1108.1030 [nucl-th].

\bibitem{Chang:2011vu}
  L.~Chang, C.~D.~Roberts and P.~C.~Tandy,
  ``Selected highlights from the study of mesons,''
  arXiv:1107.4003 [nucl-th].

\bibitem{Eichmann:2008ef}
  G.~Eichmann \emph{et~al}., 
  {Phys.\ Rev.\  C} \textbf{79}, 012202(R) (2009). 

\bibitem{Cahill:1988dx}
  R.~T.~Cahill, C.~D.~Roberts and J.~Praschifka,
  {Austral.\ J.\ Phys}.\  \textbf{42}, 129 (1989).

\bibitem{Cahill:1987qr}
  R.~T.~Cahill, C.~D.~Roberts and J.~Praschifka,
  {Phys.\ Rev.\  D} \textbf{36}, 2804(1987).

\bibitem{Roberts:2011cf}
  H.~L.~L.~Roberts, L.~Chang, I.~C.~Clo\"et and C.~D.~Roberts,
  Few Body Syst.\  {\bf 51}, 1 (2011).

\bibitem{Roberts:2011wy}
  H.~L.~L.~Roberts \emph{et al}., 
  Phys.\ Rev.\  C {\bf 83}, 065206 (2011).

\bibitem{Bender:1996bb}
  A.~Bender, C.~D.~Roberts and L.~Von Smekal,
  Phys.\ Lett.\  B {\bf 380}, 7 (1996).

\bibitem{Chang:2011ei}
  L.~Chang and C.~D.~Roberts,
  ``Tracing masses of ground-state light-quark mesons,''
  arXiv:1104.4821 [nucl-th].

\bibitem{Nakamura:2010zzi}
  K.~Nakamura {\it et al.}, 
  J.\ Phys.\ G {\bf 37}, 075021 (2010).

\bibitem{Pelaez:2006nj}
  J.~R.~Pelaez and G.~Rios,
  Phys.\ Rev.\ Lett.\  {\bf 97} (2006) 242002.

\bibitem{Buck:1992wz}
  A.~Buck, R.~Alkofer and H.~Reinhardt,
  Phys.\ Lett.\  B {\bf 286}, 29 (1992).

\bibitem{Bentz:2007zs}
  W.~Bentz \emph{et al}., 
  Prog.\ Part.\ Nucl.\ Phys.\  {\bf 61}, 238 (2008).

\bibitem{Suzuki:2009nj}
  N.~Suzuki \emph{et al}., 
  Phys.\ Rev.\ Lett.\  {\bf 104}, 042302 (2010).

\bibitem{Gasparyan:2003fp}
  A.~M.~Gasparyan, J.~Haidenbauer, C.~Hanhart and J.~Speth,
  Phys.\ Rev.\  C {\bf 68} (2003) 045207. 

\bibitem{Cloet:2008re}
  I.~C.~Clo\"et et al., 
  Few Body Syst.\  {\bf 46}, 1 (2009).

\bibitem{Roberts:1996jx}
  C.~D.~Roberts, ``Confinement, diquarks and Goldstone's theorem,''
  arXiv:nucl-th/9609039.

\bibitem{ICCloetI} I.\,C.~Cl\"oet, ``Probing the quark mass in elastic and transition form factors,'' these proceedings. 

\end{thebibliography}

\IfFileExists{\jobname.bbl}{}
 {\typeout{}
  \typeout{******************************************}
  \typeout{** Please run "bibtex \jobname" to optain}
  \typeout{** the bibliography and then re-run LaTeX}
  \typeout{** twice to fix the references!}
  \typeout{******************************************}
  \typeout{}
 }


\end{document}